\documentclass{article}
\usepackage{spconf,amsmath,graphicx}
\usepackage{amssymb,amsfonts}
\usepackage{algorithmic}
\usepackage{graphicx}
\usepackage{textcomp}
\usepackage{xcolor}
\usepackage{mathtools}
\usepackage{bm}
\usepackage{array}
\usepackage{subcaption}
\usepackage{comment}
\usepackage{hyperref}
\usepackage{enumitem}
\setlist{nosep, leftmargin=14pt}

\newcommand{\repeatthanks}{\textsuperscript{\thefootnote}}

\newcommand\copyrightnote[1]{%
  \begingroup
  \renewcommand\thefootnote{}\footnote{\kern-5pt \textcolor{white}{\rule{5pt}{2ex}}#1}%
  \addtocounter{footnote}{-1}%
  \endgroup
}

\title{``HoVer-UNet'': Accelerating HoVerNet with UNet-based multi-class nuclei segmentation via knowledge distillation}%
\name{Cristian Tommasino$^{1}$, Cristiano Russo$^{1}$, Antonio Maria Rinaldi$^{1}$\sthanks{Authors jointly supervised this work.}, Francesco Ciompi$^{2}$\repeatthanks}

\address{$^{1}$Department of Electrical Engineering and Information Technology \\University of Napoli Federico II, Napoli, Italy\\
$^{2}$Department of Pathology, Radboud University Medical Center, Nijmegen, The Netherlands\\}

\begin{document}
%
\maketitle
\begin{abstract}
We present ``HoVer-UNet,'' an approach to \emph{distill the knowledge} of the multi-branch HoVerNet framework for nuclei instance segmentation and classification in histopathology. 
We propose a compact, streamlined single UNet network with a Mix Vision Transformer backbone, and equip it with a custom loss function to optimally encode the distilled knowledge of HoVerNet, reducing computational requirements without compromising performances. 
We show that our model achieved results comparable to HoVerNet on the public PanNuke and Consep datasets with a three-fold reduction in inference time.
We make the code of our model publicly available at \small \url{https://github.com/DIAGNijmegen/HoVer-UNet}.
\end{abstract}
\section{Introduction}
\copyrightnote{This work has been submitted to the IEEE for possible publication. Copyright may be transferred without notice, after which this version may no longer be accessible.}
Nuclei panoptic segmentation, i.e., the simultaneous detection, segmentation, and classification of nuclear instances, is at the core of the automation of several tasks in digital pathology, particularly in the analysis of routine Hematoxylin and Eosin (H\&E) stained histology slides \cite{kirillov2018panoptic}.
In recent years, this task has been increasingly addressed using deep learning techniques, achieving accurate segmentation results that can power the design of computational biomarkers \cite{graham2023screening}.
However, several studies have addressed the challenge of nuclei instance segmentation and classification, with a key concern being the inference time required for accurate results \cite{hayakawa2021computational}.

\begin{figure}[!t] 
    \centering
    \includegraphics[width=1\linewidth]{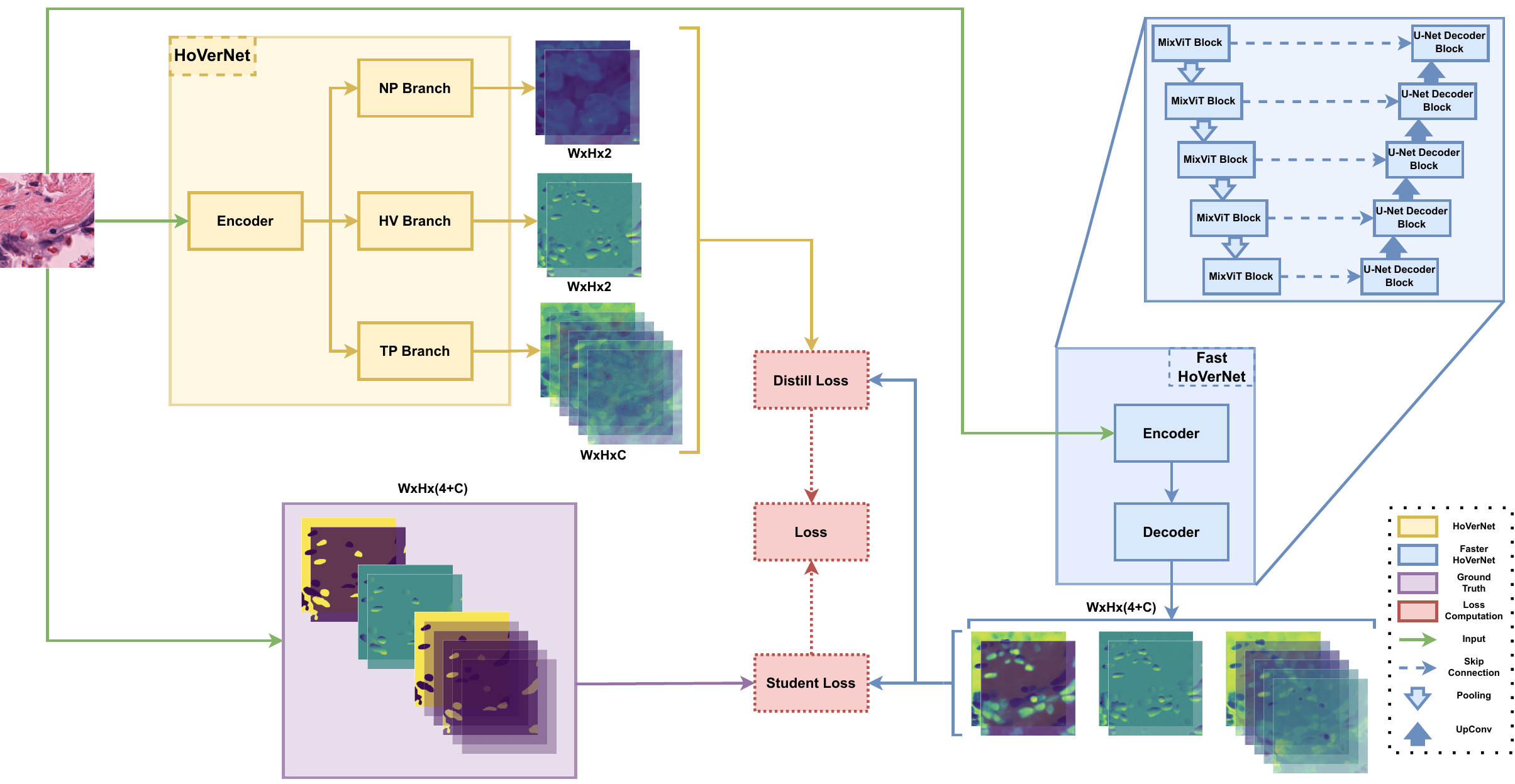}
    \caption{HoVerNet distillation framework. 
    Input image is given to the teacher (yellow component) and the student (blue component). The distillation loss is computed between the teacher output and the student output, while the student loss is computed between the generated ground truth (violet component) and student output. Lastly, both losses are combined to have the final loss (red components).}
    \label{fig:arch}
    \vspace{-0.5cm}
\end{figure}

\textbf{HoVerNet.} Common nuclei segmentation approaches involve a multi-task learning paradigm based on convolutional neural networks with a single encoder and multiple decoders, each dedicated to a specific task, followed by traditional energy-based methods such as the watershed algorithm to process output maps generated by deep learning models.
One of the best incarnations of this approach is the HoVerNet model introduced by Graham et al. \cite{graham2019hover}, which features a pre-activated ResNet50 as an encoder and three decoders, which perform nuclei prediction (NP), horizontal and vertical maps prediction (HV), and nuclei classification (NC). Afterward, HoVerNet runs post-processing using NP and HV output to obtain the nuclei instance and combines it with the TP output to classify nuclei (Figure \ref{fig:arch}). The effectiveness of this network has been shown on several public datasets \cite{jaume2021histocartography,chen2022pan}, but its dimension and complexity lead to high inference time.

\textbf{Knowledge distillation.} Introduced by Hinton et al. \cite{hinton2015distilling}, knowledge distillation enables the training of a smaller network (\emph{student}) based on a more complex network (\emph{teacher}) by using the teacher's predictions to guide and improve the student's learning process. For this, it uses a loss function that merges two hyperparameters: 1) temperature ($T$) to smooth the predictions, allowing the student to learn while reducing the influence of biases from the teacher, and 2) alpha ($\alpha$), to combine the student loss, computed between the student's predictions and the ground truth, with the distillation loss. The latter is calculated as the Kullback-Leibler divergence between the scaled softmax logits of the student and teacher, multiplied by $1/T^2$.
In this paper, we propose using KD to derive a lightweight panoptic segmentation model from the original HoVerNet, aiming to maintain HoVerNet's performance while speeding up inference time.


\section{Method}
\label{sec:3}
In this section, we present the details of HoVer-UNet, highlighting the distillation strategy utilized during the training step and introducing the proposed loss function.

\textbf{Distillation Framework.} In our KD framework, we adopt an offline technique using HoVerNet as a pre-trained teacher network. 
Given that HoVerNet performs nuclei instance segmentation and classification through three branches, our distillation strategy is based on the idea of combining all output branches of HoVerNet into a single branch network (see Figure \ref{fig:arch}.
Note that we aim to train a student that can replace only the HoVerNet backbone, not its post-processing steps, which we left unvaried.
We employ a single-branch UNet \cite{ronneberger2015u} as our student model and join all HoVerNet branch outputs into a single branch with a number of output channels equal to the total number of HoVerNet's branches. In particular, we used a Mix Vision Transformer (MixViT) \cite{xie2021segformer} as the backbone for UNet, resulting in the best combination based on our experiments.



\textbf{Loss function.} We propose a custom loss function to distill HoVerNet into a single-branch UNet. As suggested by the KD theory \cite{hinton2015distilling}, our loss is a linear combination of two losses, the \emph{student loss} between the student and the ground truth and the \emph{distillation loss} between the student and the teacher regulated by $\alpha$ parameter, and defined as:
\begin{equation}
\mathcal{L} = \alpha \cdot \mathcal{L}_{student} + (1-\alpha) \cdot \mathcal{L}_{distill}.
\end{equation}
With regards to individual loss terms, we were inspired by HoVerNet loss, and we specifically modified it for KD.
In detail, the student and distillation losses, denoted as \(\mathcal{L}_{\text{student}}\) and \(\mathcal{L}_{\text{distill}}\), respectively, are derived from a linear combination of losses across three branches: NP, HV, and TP. The total loss for each branch, \(L_{k}\), is defined as the sum of the individual losses from these branches, where \(k\) represents either the student or distillation case. 
In all formulas reported in the following, $x$ is student prediction, while $y$ for distill loss is HoVerNet prediction, and for student loss is ground truth.
For the HV branch, the loss is formulated as \(\mathcal{L}_{\text{HV}} = \text{MSE}(x,y) + \text{MSGE}(x,y)\), incorporating Mean Squared Error (MSE) and Mean Squared Gradient Error (MSGE). For the NP and TP branches, the student loss is similarly defined as \(\mathcal{L}_{\text{NP}} = \mathcal{L}_{\text{TP}} = \text{CE}(x,y) + \text{DICE}(x,y)\), combining Weighted Cross-Entropy (CE) loss with Dice loss (DICE). In contrast, the distillation loss for these branches is expressed as \(\mathcal{L}_{\text{NP}} = \mathcal{L}_{\text{TP}} = \text{CE}(x,y) + \text{KLD}(x,y)\), utilizing Weighted Cross-Entropy loss and Kullback-Leibler Divergence (KLD). 

\section{Experimental Results}
\label{sec:4}


\noindent \textbf{Datasets.} In this work we used two datasets, namely PanNuke \cite{gamper2020pannuke} for training HoVer-UNet and CoNSeP \cite{graham2019hover} for validating results on external data.
PanNuke consists of 6078 H\&E tiles, each of size $256 \times 256$, covering 19 different tissue types. It was designed for nuclei instance classification and segmentation and contains annotations of five types of nuclei: neoplastic, inflammatory, connective/soft tissue, dead, and healthy epithelial.
CoNSeP consists of 41 H\&E stained image tiles, each of size $1,000\times 1,000$ pixels at $40\times$ objective magnification, with annotations of seven types of nuclei: malignant, normal, endothelial, miscellaneous, Fibroblas, Muscle, and inflammatory.

\textbf{Metrics}. To evaluate the efficacy of our framework, we utilized the metrics as recommended in the studies of Graham et al. \cite{graham2019hover} and Gamper et al. \cite{gamper2020pannuke}. Specifically, we employed the Panoptic Quality metric, initially proposed in \cite{kirillov2018panoptic}, in conjunction with the F-score, introduced by Graham et al. \cite{graham2019hover}. 

\textbf{Training.} To train our network, we use the Adam optimizer with an initial learning rate equal to $10^{-4}$ and betas equal to $(0.99,0.9999)$. We also used a reduced on-plateau learning scheduler with a patience of 5, a scale factor of $10^{-1}$, a delta of $10^{-4}$, a minimum learning rate of $10^{-6}$, and an early stopping with the patience of 10. Furthermore, we used a depth of encoder and decoder of 5 and initialized the encoders with ImageNet-pretrained weights.

\textbf{Hyperparameter definition}. Our framework incorporates three hyperparameters: the temperature coefficient ($T$), the parameter $\alpha$, and the backbone. Throughout our experiments, we explored a spectrum of values for these parameters. For $T$, we examined values of 1, 3, and 5. For $\alpha$, we considered values of 0 and 0.5. 

\textbf{Backbone selection}. As for the backbone, we experimented with various architectures, including MixViT-B0, MixViT-B1, MixViT-B2, and MixViT-B3. We selected these models because they are substantially faster than HoVerNet and exhibit fewer Multiply-Accumulate Operations (MACs). Specifically, the most complex model has 10.63 GMac and an inference time of 0.065 seconds, compared to HoVerNet, which has 149.73 GMac and an inference time of 0.835 seconds. We computed the inference times as an average on a batch of 16 patches with shape 256x256.

\textbf{Results on PanNuke}. We evaluated several performance metrics including binary panoptic quality ($PQ_B$), multiclass panoptic quality ($PQ_M$), panoptic quality for specific classes ($PQ^{{N, I, C, D, E}}$), F-score detection ($F_d$), and F-score for individual classes ($F^{{N, I, C, D, E}}$). The categories are denoted as Neoplastic (N), Inflammatory (I), Connective/Soft Tissue (C), Dead Cells (D), and Epithelial (E).
To ensure robustness, we conducted the training using a 3-fold cross-validation (3-CV) approach, following the recommendation of the PanNuke authors \cite{gamper2020pannuke}. All reported results represent the mean across the test outcomes. Considering the extensive amount of data collected from our experiments, we present our best-performing results in comparison to DIST \cite{naylor2018segmentation}, Mask-RCNN \cite{he2017mask}, Micro-Net \cite{raza2019micro}, and HoVerNet \cite{gamper2020pannuke} \footnote{HoVerNet trained on PanNuke is fast version. \\See https://github.com/vqdang/hover\_net}. Specifically, our optimal result was achieved using the UNet model with a MixViT-B2 backbone, employing $\alpha=0.5$ and $T=1$ as the hyperparameters.

\begin{table}[!ht]
\centering
\caption{Panoptic Quality comparison on PanNuke dataset}
\label{tab:mqp}
\scalebox{0.75}{
\begin{tabular}{p{1.9cm}| *{2}{p{0.8cm}}| *{5}{p{0.8cm}}}
\hline
      \textbf{Model} & 
      $\boldsymbol{PQ_B}$ &
      $\boldsymbol{PQ_M}$ &
      $\boldsymbol{PQ^N}$ &  $\boldsymbol{PQ^I}$ &  $\boldsymbol{PQ^C}$ &   $\boldsymbol{PQ^D}$ &  $\boldsymbol{PQ^E}$ \\
\hline
    DIST &     0.5346	 &   0.3406 & 0.4390 &        0.3430 &                        0.2750 & 0.0000 &      0.2900 \\
Mask-RCNN &     0.5528 &  0.3688  &  0.4720 &        0.2900 &                        0.3000 & 0.0690 &      0.4030 \\
Micro-Net &     0.6053  &  0.4059  & 0.5040 &        0.3330 &                        0.3340 & 0.0510 &     0.4420 \\
HoVerNet &    0.6596  & 0.4629  & 0.5510 &        0.4170 &                        0.3880 & 0.1390 &      0.4910 \\
\hline
    Ours &        0.6286& 0.4475  &  0.5239 &        0.4009 &                        0.3794 & 0.0762 &      0.4779 \\
\hline
\end{tabular}}
\end{table}

Table \ref{tab:mqp} presents the PQ evaluation results. Compared with HoVerNet, our solution achieved lower scores yet in line with HoVerNet's performance, outperforming the other networks listed in the table. However, when considering the results of the $F-score$, as shown in Table \ref{tab:comp_f1}, our solution demonstrates comparable performance to HoVerNet and outperforms the other networks. Moreover, our proposed solution demonstrates a significant advantage in terms of processing speed compared to HoVerNet. This is primarily attributed to our architecture designed to reduce inference time. Additionally, we have preserved the same post-processing methodology as previously outlined, further solidifying our approach as an efficient option for this task.

    \begin{table}[!ht]
    \centering
    \caption{F-Score comparison on PanNuke dataset}
    \label{tab:comp_f1}
    \scalebox{0.95}{
    \begin{tabular}{p{1.9cm}|p{0.65cm}| *{5}{p{0.65cm}}}
    \hline
    \textbf{Model} & $\boldsymbol{F_d}$ & $\boldsymbol{F^N}$ & $\boldsymbol{F^I}$ & $\boldsymbol{F^C}$ & $\boldsymbol{F^D}$ & $\boldsymbol{F^E}$ \\
    \hline
    DIST & 0.73 & 0.50 & 0.42 & 0.39 & 0.00 & 0.35 \\
    Mask-RCNN & 0.72 & 0.59 & 0.50 & 0.42 & 0.22 & 0.52 \\
    Micro-Net & 0.80 & 0.62 & 0.52 & 0.47 & 0.19 & 0.58 \\
    HoVerNet & 0.80 & 0.62 & 0.54 & 0.49 & 0.31 & 0.56 \\
    \hline
    Ours & 0.79 & 0.64 & 0.53 & 0.48 & 0.18 & 0.62 \\
    \hline
    \end{tabular}}
\end{table}

\begin{figure}[!ht]
    \centering
    \includegraphics[width=0.99\linewidth]{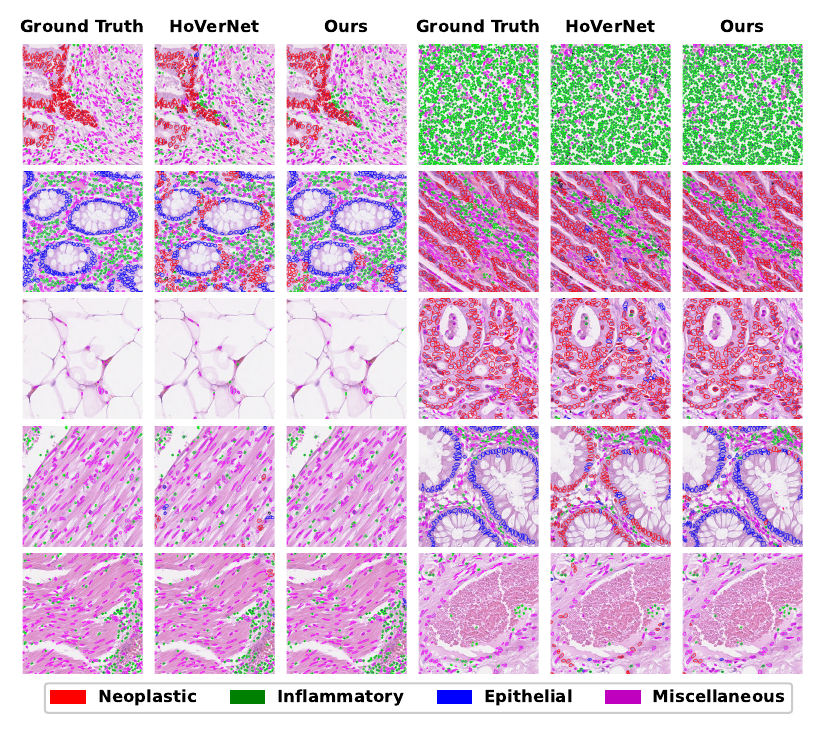}
    \caption{Nuclei segmentation and classification comparison between CoNSeP ground truth, HoVerNet, and our predictions.}
    \label{fig:ex}
\end{figure}

\textbf{Results on CoNSeP}. We undertook a comparative analysis between HoVerNet and HoVer-UNet, both pre-trained on PanNuke, leveraging the CoNSeP dataset \cite{graham2019hover} for external evaluation purposes on out-of-domain data. Given that the classification of nuclei only showed partial correspondence of classes between the PanNuke and CoNSeP and consequently between the trained HoVer-UNet and CoNSeP targets, we remapped labels into several subclasses for a more detailed comparison. The newly defined classes were the Neoplastic, Inflammatory, Epithelial, and Miscellaneous.
We mapped the Neoplastic class with \textit{PanNuke's neoplastic} and \textit{CoNSeP's dysplastic/malignant epithelial}, the Inflammatory class with \textit{PanNuke's inflammatory} and \textit{CoNSeP's inflammatory}, the Epithelial class with \textit{PanNuke's epithelial} and \textit{CoNSeP's healthy epithelial}, and the Miscellaneous class with \textit{PanNuke's dead and connective tissues} as well as CoNSeP's \emph{other types}, which include fibroblast, muscle, and endothelial tissues.
Table \ref{tab:multi_consep} shows the results demonstrating that our solution outperforms HoVerNet in terms of PQ, though it falls short in terms of F-score detection. Regarding classification metrics, our solution outperforms HoVerNet across neoplastic and epithelial nuclei; it is practically equal for miscellaneous and worse for inflammatory. Lastly, the inference time is about three times lower.
\begin{table}[!h]
\centering
\caption{Multiclass results on CoNSeP dataset}
\scalebox{0.8}{
\begin{tabular}{p{1.3cm}|p{0.65cm}| *{4}{p{0.65cm}}| p{0.65cm}| p{1.8cm}}
\hline
           \bfseries Model &  $\boldsymbol{F_d}$ & $\boldsymbol{F^N}$ & $\boldsymbol{F^I}$ &  \bfseries $\boldsymbol{F^E}$ &  $\boldsymbol{F^O}$ & \boldsymbol{$PQ$} & \textbf{Inf. Time(s)}\\
\hline
         HoVerNet & 0.818 &        0.526 &          0.758 &        0.495 &   0.559  & 0.415 & $\sim 48$\\
        Ours & 0.742 & 0.594 & 0.681  & 0.603 &0.557 & 0.599 & $\sim 17$ \\
\hline
\end{tabular}}
\label{tab:multi_consep}
\end{table} 
Figure \ref{fig:ex} shows visual examples of the results of HoVerNet and HoVer-UNet compared with the CoNSeP reference standard.
Overall, the similarity between the results supports the practical effectiveness of our approach.
\section{Conclusion}
\label{sec:5}
We have presented a novel approach for panoptic segmentation in histopathological analysis of H\&E stained images. Our method uses a KD offline approach to train a UNet network combined with a Mix Vision Transformer backbone. Our approach yields results comparable to the state-of-the-art HoVerNet model and considerably reduces inference time. Furthermore, our approach does not increase training time because our network has fewer parameters and MACs than HoVerNet, and we make HoVerNet inferences only once and then use them in the distillation loss. Lastly, considering a whole-slide image, if HoVerNet takes 1 hour, we down it to $\approx$20 minutes with HoVer-UNet, with overall results in practice comparable.
This approach performs similarly to leading models and lays the groundwork for the progression of swifter computational pathology solutions, such as biomarkers and automated quantification for subsequent tasks. Notably, our method is approximately three times faster than HoVerNet. We have made our solution publicly available to promote its use in future developments in computational pathology.
\section{Compliance with ethical standards}
\label{sec:ethics}
This research study was conducted retrospectively using human subject data available in open access by the Tissue Image Analytics (TIA) Centre. Ethical approval was not required, as confirmed by the license attached with the open-access data.
\section{Acknowledgments}
\label{sec:acknowledgments}
This work was partly supported by a research grant from the Netherlands Organization for Scientific Research (project number 18388).
F.C. was Chair of the Scientific and Medical Advisory Board of TRIBVN Healthcare, France, and received advisory board fees from TRIBVN Healthcare, France in the last five years. 
He is shareholder of Aiosyn BV, the Netherlands. We also acknowledge financial support from the project Digital Health Solutions in Community Medicine (DHEAL-COM) E63C22003790001 PNC-E3-2022-23683267 PNC - HLS-DH.

\bibliographystyle{IEEEbib}

\clearpage
\end{document}